\documentstyle[12pt,aaspp4]{article}
\begin{document}
\title{On the Origin of Narrow, Very Long, Straight
Jets from Some Newly Forming Stars}

\author{Howard D. Greyber}

\affil{10123 Falls Road, Potomac, MD 20854, U.S.A.}

\begin{abstract}

Observations have shown the existence of narrow, very long,
straight jets emitted by some newly forming stars (1).
It is highly likely that stars forming in the plane of a spiral
galaxy do so in the presence of an almost uniform   magnetic
field. In the Strong Magnetic Field model (SMF), gravitational
collapse of a highly conducting plasma in the presence of such
a field will result in the formation of a stable, highly relativistic current
loop (storage ring) around the central object.
The concept was first described by Greyber (2-14).  In the figures
in Mestel \& Strittmatter (15), one can see such a storage ring
beginning to form. Such an increasing dipole magnetic field
(formed temporarily for 10$^4$  to 10$^6$  years)
will produce, accelerate and confine a narrow, very long, straight jet.
When the density becomes too high, either the loop is destroyed,
or the current-carrying plasma ring is buried inside the newly
forming star and is the source of primordial stellar magnetism.

\end{abstract}

\newpage

\section{Introduction}

The famed physicist Enrico Fermi introduced equipartition
into astrophysics in the Forties, and did it right. He discussed shock
waves, and obviously turbulence close to the shock made equipartition
a reasonable assumption there.  However astronomers soon
forgot the caveat, i.e. that there was no reason to assume plasma
turbulence producing equipartition must exist everywhere in
astrophysics. Thus, the very small magnetic fields, often deduced
and published, from assuming equipartition, are irrelevant in many situations.
Actually equipartition applies only when the
physics demands it does!{italics?}

Thus, for almost four decades, there has been a
widespread misconception that "equipartition" between the particle
energy and the magnetic field energy was absolutely necessary in
most astrophysical situations. Very strong cosmic magnetic fields
are accepted as real in white dwarfs and neutron
stars. However until recently it has been alleged that strong
magnetic fields could not exist in the cores of quasars and active
galactic nuclei (AGN), nor in newly forming stars.

For 35 years, the Strong Magnetic Field model,
Greyber (2-14), has argued forcefully for what Chi and Wolfendale (16)
wrote recently. "there is, however, no compelling justification for
this assumption of equipartition". Originally created to explain
spiral arms and answer Oort's famous questions, SMF has since been
applied to the physical model of the central engine of AGNs,
jet formation, galactic energetics and morphology, gamma ray bursts, etc.

\section{The Strong Magnetic Field Model (SMF)}

Greyber (17) has proposed an original model,
within the Big Bang hypothesis, involving cosmical magnetism
as well as gravitation, that explains the origin of large-scale
primordial magnetic fields, as well as the origin of the observed highly
 structured nature (thin sheets of galaxies and voids) of matter
in the Universe.
Note that Pietronero et al (18) conclude, "galaxy correlations
are fractal and not homogeneous up to the limits of  the available catalogues".

Kosowsky and Loeb (19) recently emphasized
"The origin of the primordial field is still a subject of
speculation.  In the past, various indirect theoretical arguments
were used to favor the dynamo amplification mechanism over the primordial
origin alternative.
However recent studies argue that a galactic dynamo should
saturate due to the rapid growth of a fluctuating small-scale
field before it can actually result in a coherent large-scale
field of the type observed in galactic disks. The view that the
galactic field may, in fact, be primordial gains additional support
from observations of damped Lyman     absorption systems in QSO
spectra at  Z      2.  The potential existence of a
primordial magnetic field is also consistent with observations of
clusters of galaxies. Faraday  rotation measurements of radio sources
inside and behind clusters indicate strong magnetic fields in many of them."

When one considers a galaxy or quasar  forming by
gravitational collapse of a giant, highly conducting, plasma
cloud containing an almost uniform primordial magnetic field,
SMF argues that a new physical construct, a storage ring, is created.
Since the topology is very similar, the very same process occurs during the
formation of a star under gravity. The gravitationally bound current
loop, or
storage ring, is extremely intense and highly relativistic.
The bursting force of this very strong unified magnetic field system
is in equilibrium, balancing the gravitational force between the
slender toroidal plasma (bound to the current loop in
side the very slender toroid by the Maxwell "frozen-field" condition),
and the central massive object.

The morphology and energetics of objects of galactic
dimension are determined in SMF by the ratio of magnetic field
energy to rotational energy in that particular object. The ratio
is extremely high for quasars and blazars, and decreases steadily for giant
 elliptical and radio galaxies, Seyferts, Markarians, is low in
 ordinary spirals and is close to zero for the ordinary elliptical
 galaxies. However it is important to note that the AGN activity
we now observe is a function of the accretion of
matter into the central engine of the object.

A diagram of the SMF central engine for AGN is in
references (10) and (14). It is the same for around a newly
forming star, except that the central mass concentration is
the protostar. High energy particles in this completely coherent,
relativistic current
 loop store a significant fraction of the huge energy of
 gravitational collapse.  The dipole magnetic field bound
 to the protostar probably contributes to the loss of angular
 momentum of the contracting protostar.

One can see such a storage ring forming in the
figures in a brilliant, pioneering paper by Mestel and Strittmatter (15).
They analyzed the effect of Ohmic diffusion on the magnetic field
distribution of a gravitationally bound magnetic gas cloud, illustrating
how the magnetic field topology changes as the cloud
field detaches from the background field.

A storage ring, once formed, is uniquely stable.
Due to coherence, one part of the loop does not radiate in the magnetic
field of another part.  However, if a fluctuation or "bump" occurs
somewhere along the loop, the electro
ns in the "bump" will suddenly radiate furiously in the immensely strong local
magnetic field, the energy in the fluctuation will dissipate
rapidly, and the storage ring will return quickly to its undisturbed
configuration.  A relevant point is that the largest external
 perturbation to a storage ring is limited to solid matter,
i.e. objects not much larger than the planet Jupiter, since
stars, made of plasma, would break up before penetrating close
to the intense magnetic field of the loop.

Klein and Brueckner (20) investigated the motion of a
plasma under the action of an increasing magnetic field from a
stationary coil. They found that the efficiency of conversion
of stored energy into kinetic energy was about 5-10\%.  A simil
ar or greater efficiency is then expected in SMF for the expulsion of the
extremely high conductivity plasma from around a newly forming
star, forming a jet.  The increasing dipole magnetic field, as
the contraction under gravity forming the star continues, accelerates
and keeps the plasma jet narrow and confined for extremely long
distances from the protostar.

The SMF model predicts that jets are formed by
the successive emission of  blobs of plasma from the AGN or
stellar "central engine".  Jets composed of blobs of plasama
are just what is observed, both in jets from galaxies and quasars,
and also observed as
well in some newly forming stars like HH-30, which has been observed
recently with HST.

\section{Conclusion}

The fact that jets are not well defined in many newly
forming stars is understandable considering the relatively high
density of the plasma in the vicinity of the protostar, and the
disruption to the storage ring that may occur from binary o
r multiple prot
ostars forming in close proximity.  When the plasma density
becomes too high, either the storage ring is destroyed, or,
in some instances, the current-carrying plasma loop is buried
inside the newly forming star and is the source of primordial
stellar magnetism.

Evidence for the storage ring of current around
some newly forming stars, with narrow, straight, long jets,
hopefully will be found as observational resolution improves.
This would be very important because it would validate the same
topology as applied to
 the physics of the central engine of quasars and galaxies.
 So far, the SMF model appears to fit the observations.

\section{References}

\noindent
1. DeYoung, D., 1991, Science {\bf 252}, 389.\\

\noindent
2. Greyber, H. D., 1961, Trans. of the I.A.U. XIB, 332; Report of Commission
33.\\

\noindent
3. Greyber, H. D., 1962, U.S.A.F.O.S.R. Research Report No. 2958,
"On the Steady State Dynamics of Spiral Galaxies".\\

\noindent
4. Greyber, H. D., 1963, Astron. J. {\bf 68}, 536.
\noindent

5. Greyber, H. D., 1964, Chapter 31, in
"Quasistellar Sources and Gravitational Collapse". ed. Ivor Robinson et al,
University of Chicago Press (First Texas Symposium on Relativistic
Astrophysics).\\

\noindent
6. Greyber, H. D., 1967, in "Instabilitie Gravitationelle et Formation des
Etoiles,
des Galaxies, de Leurs Structures Caracteristique",
Memoirs Royal Society of Sciences of Liege, XV, 189-196.\\

\noindent
7. Greyber, H. D., 1967, Publications Astron. Soc. of the Pacific, {\bf 79},
341.\\

\noindent
8. Greyber, H. D., 1984, 11th Texas Symposium,
Annals of the New York Acad. of Sciences {\bf 422}, 353.\\

\noindent
9. Greyber, H. D., 1988 in "Supermassive Black Holes",
 ed. M. Kafatos, Kluwer Acad. Press, 360.\\

\noindent
10. Greyber, H. D., 1989, Comments on Astrophysics {\bf 13}, 201.\\

\noindent
11. Greyber, H. D., 1989, in "The Center of the Galaxy", ed. Mark Morris,
Kluwer Acad. Press, 335.\\

\noindent
12. Greyber, H. D., 1990, 14th Texas Symposium, Annals of the New York Acad. of
Sciences, 571, 239.\\

\noindent
13. Greyber, H. D., 1993, in "Compton Gamma Ray Observatory", A.I.P. Conference
Proceedings 280, 569.\\

\noindent
14. Greyber, H. D., 1994, in "COSMICAL MAGNETISM Contributed Papers",
edited by D. Lynden-Bell, Institute of Astronomy, Cambridge, England,
110-118, NATO Advanced Study Institute, in honour of Prof. L. Mestel, FRS.\\

\noindent
15. Mestel, L. \& Strittmatter, P., 1967, M.N.R.A.S. {\bf 137}, 95.\\

\noindent
16. Chi, X. \& Wolfendale, A. W., 1993, Nature {\bf 362}, 610.\\

\noindent
17. Greyber, H. D., 1996, in "Clusters, Lensing and the Future of the
Universe",
A.S.P. Conf. Series Vol. 88, 298.\\

\noindent
18. Pietronero, L. et al, 1997, to be published in
"Critical Dialogues in Cosmology", World Scientific, editor N. G. Turok.\\

\noindent
19. Kosowsky A. \& Loeb, A., 1996, Ap.J. {\bf 469},  1.\\

\noindent
20. Klein, M. M. \& Brueckner, K. A., 1960, Journal  of Applied Physics {\bf
31}, 1437.\\

\end{document}